\pdfoutput=1
\documentclass[11pt,a4paper]{article}
\usepackage{jheppub,color,amsfonts,mathrsfs,amsthm}
\usepackage[caption=false]{subfig}
\usepackage[titletoc,title]{appendix}

\newcommand{\beq}{\begin{eqnarray}}
\newcommand{\eeq}{\end{eqnarray}}
\newcommand{\non}{\nonumber\\}

\newcommand{\p}{\partial}

\newcommand{\U}{\qopname\relax o{U}}
\newcommand{\SU}{\qopname\relax o{SU}}

\renewcommand{\i}{\mathrm{i}}
\renewcommand{\d}{\mathrm{d}}

\newcommand{\calF}{\mathcal{F}}

\newcommand{\calA}{\mathcal{A}}
\newcommand{\calR}{\mathcal{R}}
\renewcommand{\S}{\mathbb{S}}

\newcommand{\R}{\mathbb{R}}

\newcommand{\C}{\mathbb{C}}
\newcommand{\Z}{\mathbb{Z}}
\renewcommand{\H}{\mathbb{H}}

\title{Magnetic Impurities, Integrable Vortices and the Toda Equation} 

\author{Sven Bjarke Gudnason$^1$,}
\affiliation{$^1$Institute of Contemporary Mathematics, School of
  Mathematics and Statistics, Henan University, Kaifeng, Henan 475004,
  P.~R.~China}
\emailAdd{gudnason(at)henu.edu.cn}
\author{Calum Ross$^{2,3}$}
\affiliation{
$^2$Department of Physics, University College Cork, Cork T12
  K8AF, Ireland\\
$^3$Department of Physics and Research and Education Center for Natural Sciences, Keio \\ University, Hiyoshi 4-1-1, Yokohama, Kanagawa 223-8521, Japan}
\emailAdd{c.ross(at)keio.jp}

\abstract{
The five integrable vortex equations, recently studied by Manton, are
generalized to include magnetic impurities of the Tong-Wong type.
Under certain conditions these generalizations remain integrable. We
further set up a gauge theory with a product gauge group, two complex
scalar fields and a general charge matrix. The second species of
vortices, when frozen, are interpreted as the magnetic impurity for
all five vortex equations. We then give a geometric compatibility
condition, which enables us to remove the constant term in all the 
equations. This is similar to the reduction from the Taubes equation to
the Liouville equation. We further find a family of charge matrices
that turn the five vortex equations into either the Toda equation or
the Toda equation with the opposite sign. We find exact analytic
solutions in all cases and the solution with the opposite sign appears
to be new.
}

\begin{document}
\maketitle

\section{Introduction}

Vortex equations have been studied in many fields of physics ever
since they made their appearance in the Ginzburg-Landau equations
describing magnetic fluxes penetrating a type-II superconductor,
giving rise to the Abrikosov lattice, see
e.g.~\cite{Taubes:1979tm,MS,JT,YangsBook}. 
Other vortex equations have been discovered, like the Jackiw-Pi
vortices in nonrelativistic Chern-Simons
theory \cite{Jackiw:1990tz,Jackiw:1990mb}, the Ambj\o rn-Olesen
vortices in the electroweak theory \cite{Ambjorn:1988fx} and Popov
vortices in the reduction a l\'a Witten \cite{Witten1} of self-dual Yang-Mills theory to the
2-sphere \cite{Popov} instead of the hyperbolic plane \cite{Witten1}. 
All of the above vortex equations were put on equal footing by Manton
in Ref.~\cite{Manton2} and an extra equation was included, dubbed the
Bradlow vortex equation because of its resemblance to the Taubes
vortex equation in the Bradlow limit. 
These ``exotic'' vortex equations are all integrable in two dimensions, but not all on
the same base Riemann surface.
It turns out that there is a direct connection between the signs (or
``0'') of the Fayet-Iliopoulos parameter (the constant term) of the
vortex equation and the (constant) curvature of the Riemann surface on
which they are integrable. 
The coefficient in front of the vortex field has the same property,
not related to the base Riemann surface, but to the constant curvature of
the Baptista surface \cite{Baptista}.
Witten showed in the seminal paper that the integrable version of the
Taubes equation, which on the hyperbolic plane becomes the Liouville
equation, has the interpretation as a suitable reduction of integrable
instantons in $\R^4$ to a sphere and a hyperbolic
plane \cite{Witten1}. 
Doing the opposite reduction, one arrives at the Popov
vortex equations, which are integrable on a 2-sphere \cite{Popov}. 
Finally, such a reduction was generalized to all the exotic vortex
equations by Contatto and Dunajski \cite{CD}. 

Impurities have recently become an active area of theoretical
research, perhaps starting with the elegant interpretation given to
magnetic impurities in the Taubes equation by Tong and Wong, as a
second ``frozen'' vortex species \cite{TW}.
Impurities in integrable hyperbolic (Taubes) vortices have been
studied by Cockburn-Krusch-Muhamed \cite{CKM}, where the interpretation
of a delta-function impurity as a source of extra vortices is made.
Impurities have further made possible the discovery of spectral walls
in the dynamics of kinks in $1+1$ dimensions \cite{Adam:2018tnv}. 

In this paper we generalize all five exotic vortex equations to
include an impurity field, and discuss when they remain
integrable.
Then we consider a $\U(1)^2$ gauge theory with two complex scalar
(Higgs) fields and a general charge matrix, giving rise to a
nontrivial mixing of the vortex equations, following
Ref.~\cite{GEN2020}. We generalize this theory to account for all five
vortex equations.
We find a geometric compatibility condition, for which we can remove
the Fayet-Iliopoulos (constant) term from the entire system of
equations.
This puts a constraint on the charge matrix as well as the
Fayet-Iliopoulos (constant) parameters.
Choosing a special family of charge matrices, we reduce the system
of equations to the integrable Toda system, for the case of a specific
sign in the vortex equations.
We further generalize the solution of
Kostant-Leznov-Saveliev
(KLS) \cite{Kostant:1979qu,Leznov:1979td,Leznov:1979yx} 
to both signs of the constant and explain the trivial case of
coupled Bradlow equations.
This establishes exact analytic solutions, in closed form, for the
Taubes-Toda, Popov-Toda, Jackiw-Pi-Toda, Ambj\o rn-Olesen-Toda and
Bradlow-Toda vortices.
The Taubes-Toda equation has the opposite sign to the normal
Toda equation and the KLS solution is not directly applicable.
We find a generalization of the KLS solution to the Toda equation of
the opposite sign, which to the best of our knowledge is new.
Interestingly, the normal Toda equation corresponds naturally to the
coupled Popov, coupled Jackiw-Pi and coupled Ambj\o rn-Olesen
equations on certain background manifolds.
Finally, we generalize the frozen vortex explanation for impurities of
Tong and Wong to all five exotic vortex equations.

The paper is organized as follows.
In Sec.~\ref{sec:exotic_review}, we review the five exotic vortex
equations. In Sec.~\ref{sec:mag_imp}, we generalize the exotic vortex
equations to include magnetic impurities. In Sec.~\ref{sec:product},
we set up the gauge theory with product gauge groups for exotic
vortices with a generic charge matrix and find the geometric
compatibility condition. In Sec.~\ref{sec:Toda} we find the solution
for the charge matrix that reduces the coupled vortex equations to
integrable Toda equations. In Sec.~\ref{sec:impurity_from_product}, we
use the theory to rederive the magnetic impurity as a frozen second
species of vortex. In Sec.~\ref{sec:general_coupled}, we consider some
examples of nonintegrable vortex equations and give an example of a
numerical solution. Finally, we conclude the paper with a discussion
in Sec.~\ref{sec:discussion}.
Before starting off, we will quickly set the conventions used in the
rest of the paper.

\subsection{Conventions for local geometry}

It will prove convenient to set the notation and conventions for the
paper.
For a Riemann surface $M_0$ with constant Gauss curvature $K_0$ we
work in terms of a local complex coordinate, $z$. The Riemann
surface\footnote{Here $M_0$ will either be $\S^2$, $\H^2$ or
$\R^2$. In the two latter cases $z$ is a global coordinate, whereas
for $\S^2$ we need at least two patches to cover the space
globally.} 
has the metric 
\begin{equation}
\d s^2 = \Omega_0\,\d z\d\bar{z}
= \frac{4}{\left(1+K_0|z|^2\right)^2}\,\d z\d\bar{z},
\end{equation}
which admits the local complexified frame
\begin{equation}
e = \frac{2\d z}{1+K_0|z|^2}.
\end{equation}
In terms of the frame field, the structure and Gauss equations are
\begin{align}
\d e -\i e\wedge \Gamma &=0,\\
\d\Gamma = \calR &= \frac{\i}{2}K_0\, e\wedge \bar{e},
\end{align}
where
\beq
\Gamma=\i K_0\frac{z\d\bar{z}-\bar{z}\d z}{1+K_0|z|^2},
\eeq
is the spin connection and $\calR$ is the curvature 2-form.

\section{Exotic vortex equations}\label{sec:exotic_review}

In Ref.~\cite{Manton2} a generalization of the standard hyperbolic
vortex equations was introduced. These vortices are pairs $(\phi,A)$
of a connection $A$ and a smooth section $\phi$ on a line bundle over
a Riemann surface, $M_0$. In the case that $M_0$ is non-compact, appropriate
asymptotics, $|\phi|\to 1$ on $\partial M_0$, need to be imposed so
that the energy is finite \cite{MS,JT}.
These exotic vortex equations are 
\beq
(\d\phi-\i A\phi)\wedge e = 0, \qquad
F = \d A = \left(\lambda_0-\lambda|\phi|^2\right)\,\omega_0, \label{eq:vtx}
\eeq
where $\omega_0$ is the K\"ahler form on $M_0$.

In terms of a local complex coordinate $z$ on $M_0$ these become
\begin{align}
\partial_{\bar{z}}\phi - \i A_{\bar{z}}\phi &= 0,\\
 F_{z\bar{z}}=\partial_{z}A_{\bar{z}}-\partial_{\bar{z}}A_{z}&=
 \frac{\i}{2}\Omega_0\left(\lambda_0-\lambda|\phi|^2\right).
\end{align}

By decomposing the Higgs field as $\phi=e^{h+i\chi}$ a
generalization of the Taubes equation is arrived at
\begin{equation}
-\frac4{\Omega_0}\partial_{z}\partial_{\bar{z}} h
= \big(\lambda_0-\lambda e^{2h}\big)
-\frac{2\pi}{\Omega_0}\sum_{r=1}^N\delta(z-Z_r),
\label{eq:gen_taubes}
\end{equation}
where $\Omega_0$ is the conformal factor relating the metric on $M_0$
to the flat metric on $\R^2$ and $Z_r\in\C$ are the positions of $N$,
not necessarily separated, vortices.

A scaling argument given in Ref.~\cite{Manton2} allows us to restrict
$\lambda_0$ and $\lambda$ to take the values $-1,0,1$ giving nine
possible equations.  
By integrating this equation, and imposing that the left-hand side be
positive, as it is the integral of the magnetic field corresponding to
the magnetic flux, allows us to exclude four cases.
This leaves the five cases:
\begin{center}
\begin{tabular}{rlrr}
&   Vortices & $\lambda_0$ & $\lambda$\\\hline
1.& Hyperbolic vortices & $1$ & $1$\\
2.& Popov vortices & $-1$ & $-1$\\
3.& Jackiw-Pi vortices & $0$ & $-1$\\
4.& Ambj\o rn-Olesen vortices & $1$ & $-1$\\
5.& Bradlow vortices & $1$ & $0$
\end{tabular}
\end{center}
We refer to these as the $(\lambda_0,\lambda)$ vortex equations.
As was observed in Ref.~\cite{Manton2} and we will review below, these
equations are integrable on a constant curvature Riemann surface with
Gauss curvature $K_0=-\lambda_0$.

Hyperbolic vortices have been well studied, many of the details are
given in Ref.~\cite{MS}, and are integrable on 2-dimensional
hyperbolic space, $\H^2$. Most commonly the Poincar\'e disk model of
$\H^2$ (here we take the unit disk) is used. On the disk a hyperbolic
vortex is given by a holomorphic map 
\begin{equation}
f : \H^2\to \H^2,
\end{equation}
such that $|f|\to 1$ as $|z|\to 1$. This means that $f$ can be
represented as a finite Blaschke product as was first established
in the seminal paper \cite{Witten1}. 

The Popov vortex equations were introduced in Ref.~\cite{Popov} as a
dimensional reduction of $\SU(1,1)$ instantons. These equations on
$\S^2$ were further considered in Ref.~\cite{Manton1} where solutions
were expressed in terms of rational maps 
\begin{equation}
R : \S^2\to \S^2.
\end{equation}

Bradlow vortex equations, first introduced in Ref.~\cite{Manton2}
have exact solutions on $\H^2$, but due to the
simplicity of the vortex equation it does in fact have integrable
solutions on many compact domains of various
curvatures, including vanishing curvature \cite{Gudnason:2017jsn}. 

More generally, we call the $(\lambda_0,\lambda)$ vortex equations
integrable when they can be solved by a holomorphic map between two
constant curvature Riemann surfaces,  
\begin{equation}
f : M_0\to M,
\end{equation}
where $M_0$ has Gauss curvature $K_0=-\lambda_0$ and $M$ has Gauss
curvature $K=-\lambda$ (away from ramification points of $f$ that
correspond to the zeros of $\phi$). In fact, a general expression for
the norm-squared of the Higgs field is \cite{Manton2}
\begin{equation}
|\phi|^2=\frac{\left(1-\lambda_0|z|^2\right)^2}{\left(1-\lambda|f|^2\right)^2}
\left\lvert\frac{\d f}{\d z}\right\rvert^2,
\end{equation}
where $z$ is a local complex coordinate on $M_0$.
Notice that the norm of the local section (Higgs field) is given by a
Jacobian of the coordinate transformation between the Riemann surfaces  $M_0$
and $M$, multiplied by the holomorphic tangent of the function
$f$. 
Vortices are thus ramification points of the auxiliary map $f$.
It is also known from Ref.~\cite{CD} that all these exotic vortex
equations can be reached by dimensional reduction of $G$-instantons
from the four manifold $M_0\times\Sigma$ for an appropriate choice of
gauge group $G$ and Riemann surface $\Sigma$.
For the well known hyperbolic case $G=\SU(2)$ and $\Sigma=S^2$
while for Popov vortices $G=\SU(1,1)$ and $\Sigma=\H^2$.

Following Ref.~\cite{Manton2} the potential energy piece of the
Ginzburg-Landau action functional, adapted to the
$(\lambda_0,\lambda)$ vortex, is given by
\begin{align}
V=\frac12\int_{M_0}\left(\frac{1}{\Omega_0^2}(F_{12})^2
+\frac{2\lambda}{\Omega_0}|D_i\phi|^2
+\left(\lambda_0-\lambda|\phi|^2\right)^2\right) \Omega_0\;\d^2x,\qquad
i=1,2,
\end{align}
where $F_{12}=-2\i F_{z\bar{z}}$ and
\beq
\d\phi - \i A\phi = D_i\phi\,\d x^i, \qquad i=1,2.
\eeq
A Bogomol'nyi type argument on this functional gives rise to the
vortex equations \eqref{eq:vtx},
\begin{align}
V=\frac12\int_{M_0}\left[
\left(\frac{F_{12}}{\Omega_0} - \lambda_0 + \lambda|\phi|^2\right)^2
+\frac{8\lambda}{\Omega_0}|D_{\bar{z}}\phi|^2\right] \Omega_0\;\d^2x
+2\pi\lambda_0 k,
\end{align}
where we have dropped a boundary term and defined the magnetic flux
\beq
k \mathrel{\mathop:}= \frac{1}{2\pi}\int_{M_0} F
= \frac{1}{2\pi}\int_{M_0}F_{12}\;\d^2x.
\eeq
The Bogomol'nyi bound is then
\beq
V \geq V_{\rm BPS} = 2\pi\lambda_0 k.
\eeq
where BPS stands for Bogomol'nyi-Prasad-Sommerfield.

\section{Magnetic impurities}\label{sec:mag_imp}

Magnetic impurities are an important but also a simple type of
impurity found in materials, like superconductors.
In Ref.~\cite{TW} vortices in the presence of a magnetic impurity,
$\sigma(x)$, were considered and the latter is taken into account by
simply changing the potential term as
\begin{equation}
\frac12\left(\lambda_0-\lambda|\phi|^2\right)^2\qquad\longrightarrow\qquad
\frac12\left(\lambda_0-\lambda|\phi|^2-\sigma(x)\right)^2,
\end{equation}
as well introducing a static source for the magnetic field
\begin{equation}
L_{\text{impurity}}=2\int_{M_0}\sigma(x)F_{12}\;\d^2x.
\end{equation}
Note that the magnetic field, $B$, is given by
$B=\Omega_0^{-1}F_{12}$ and hence the conformal factor $\Omega_0$
cancels out in the source term for the impurity.

A Bogomol'nyi argument still works and results now in the first-order
vortex equations 
\begin{equation}
(\d\phi - \i A\phi)\wedge e = 0, \qquad
F = \d A = \left(\lambda_0 - \lambda|\phi|^2 - \sigma(x)\right)\,\omega_0,
\end{equation}
or in terms of local coordinates
\begin{align}
\p_{\bar{z}}\phi - \i A_{\bar{z}}\phi &= 0, \label{eq:selfdual}\\
F_{z\bar{z}} = \p_z A_{\bar{z}} - \p_{\bar{z}} A_z
&= \frac{\i}{2}\Omega_0\left(\lambda_0 - \lambda|\phi|^2 - \sigma(z,\bar{z})\right).
\label{eq:imp_vtx}
\end{align}

In Ref.~\cite{CKM}, the case of Taubes vortices,
$(\lambda_0,\lambda)=(1,1)$, was explored and the effect of the impurity
on the moduli space metric was considered.
In particular, if the integrable case of Taubes vortices on the
hyperbolic disk is taken, then for a delta function defect the vortex
equations are still integrable.
We shall see shortly that this is also the case for the other
integrable $(\lambda_0,\lambda)$ vortex equations.

\subsection{Exotic Liouville-type equations with impurities}\label{sec:exotic_liouville}

Applying the standard decomposition of $\phi=e^{h+\i\chi}$ and
using the first vortex equation \eqref{eq:selfdual} to solve for $A$
we can reduce Eqs.~\eqref{eq:selfdual}-\eqref{eq:imp_vtx} to a single
equation for $h$ of Liouville-type: 
\begin{equation}
-\frac{4}{\Omega_0}\partial_{z}\partial_{\bar{z}}h
=\big(\lambda_0-\lambda e^{2h}-\sigma(z,\bar{z})\big)
-\frac{2\pi}{\Omega_0}\sum_{r=1}^{N}\delta(z-Z_{r}),
\end{equation}
where the delta function sources are at the points $Z_r$ and
correspond to the zeros of $\phi$.

We take the conformal factor to be
\begin{equation}
\Omega_0=\frac{4}{\left(1-\lambda_0|z|^2\right)^2},
\end{equation}
that is, $M_0$ is a Riemann surface with constant curvature
$K_0=-\lambda_0$. For the case of $\H^2$ and $\S^2$ this conformal
factor corresponds to a disk of radius $1$ and a sphere of radius $1$,
respectively.

Following the approach in Ref.~\cite{Witten1}, we make the substitution 
\begin{equation}
h = g + \log\left(\frac{1}2\left(1-\lambda_0|z|^2\right)\right),
\end{equation}
such that
\begin{equation}
4\partial_{z}\partial_{\bar{z}}h
= 4\partial_{z}\partial_{\bar{z}}g-\lambda_0\Omega_0.
\end{equation}
This transforms the equation for $h$ into
\begin{equation}
4\partial_{z}\partial_{\bar{z}}g
=\lambda e^{2g}+\Omega_0\sigma+2\pi\sum_{r=1}^{N}\delta(z-Z_{r}),
\label{eq:gen_Taubes}
\end{equation}
with $K=-\lambda$ being the Gauss curvature of $M$ (away from the
points $Z_r$ which correspond to the zeros of $\phi$).

There are three cases where this reduces to Liouville's equation.
\begin{itemize}
\item The first is when the defect is zero, $\sigma=0$, as then 
Eq.~\eqref{eq:gen_Taubes} becomes \cite{Manton2}
\begin{equation}
4\partial_{z}\partial_{\bar{z}}g
= \lambda e^{2g} + 2\pi\sum_{r=1}^{N}\delta(z-Z_{r}),
\label{eq:pure_gen_Taubes}
\end{equation}
which has the solution
\beq
g = -\log\frac{1-\lambda|f|^2}{2} + \frac12\log\left|\frac{\d f}{\d z}\right|^2,
\label{eq:gsol}
\eeq
and leads to \cite{Manton2}
\begin{equation}
\phi=\frac{1-\lambda_0|z|^2}{1-\lambda|f|^2}\frac{\d f}{\d z}, \qquad
A_{\bar{z}}=-\i\partial_{\bar{z}}\log\left(\frac{1-\lambda_0|z|^2}{1-\lambda|f|^2}\right),
\end{equation}
in the unitary gauge.
\item The second case is when the defect is given in terms of delta
functions
\beq
\sigma=\frac{2\pi}{\Omega_0} \sum_{j=1}^{K}\alpha_j\delta(z-Z_{N+j}), \qquad
\alpha_j\in\mathbb{R}_{>0},
\eeq
as then Eq.~\eqref{eq:gen_Taubes} becomes
\begin{equation}
4\partial_{z}\partial_{\bar{z}}g
= \lambda e^{2g}
+2\pi\sum_{j=1}^{K}\alpha_j\delta(z-Z_{N+j})
+2\pi\sum_{r=1}^{N}\delta(z-Z_{r}).
\label{eq:extra_sources1_Taubes}
\end{equation}
Only when $\alpha_j\in\Z_{>0}$ are positive integers, the impurities have the
interpretation of being sources for vortices.
If $\alpha_j\mathrel{\mathop:}=1$, $\forall j$, we have
\beq
4\partial_{z}\partial_{\bar{z}}g
= \lambda e^{2g}
+ 2\pi\sum_{r=1}^{N+K}\delta(z-Z_{r}),
\label{eq:extra_sources2_Taubes}
\eeq
which is still completely general because $\{Z_r\}$ are allowed to be 
coincident.

In Ref.~\cite{CKM} it was pointed out that the defect being a delta
function does not make sense in the Lagrangian where a square of a
delta function would appear, but the Eqs.~\eqref{eq:imp_vtx}
and \eqref{eq:gen_Taubes} are still sensible to consider.
It is also important to note that we want $\phi$ to be an $N$ vortex
solution however, Eq.~\eqref{eq:extra_sources2_Taubes} would naturally 
lead to an $N+K$ vortex solution.

Consider the case $K=1$, $\alpha\mathrel{\mathop:}=\alpha_1\in\R_{>0}$ a
positive real constant but not necessarily an integer,
set $Z_{N+1}\mathrel{\mathop:}=0$, and denote by
$f(z)\mathrel{\mathop:}=z\tilde{f}(z)$ a holomorphic function with $N$ ramification
points, which enters the solution $g$ of Eq.~\eqref{eq:gsol} to the
impurity-less Taubes Eq.~\eqref{eq:pure_gen_Taubes}.
Then we have that $f(z)\mathrel{\mathop:}=z^{\alpha+1}\tilde{f}(z)$, plugged into $g$
of Eq.~\eqref{eq:gsol}, is a solution to the Taubes equation with the
impurity (and general $\alpha\in\R_{>0}$)
\eqref{eq:extra_sources1_Taubes}.
The Higgs field corresponding to $g$ with
$f(z)\mathrel{\mathop:}=z^{\alpha+1}\tilde{f}(z)$ reads
\begin{equation}
\phi=\frac{1-\lambda_0|z|^2}{1-\lambda|z|^{2\alpha+2}|\tilde{f}|^2}
\left(\left(\alpha+1\right)z^{\alpha}\tilde{f}+z^{\alpha+1}\frac{\d\tilde{f}}{\d z}\right),
\end{equation}
in a certain choice of gauge.
As pointed out in Ref.~\cite{CKM} for hyperbolic vortices
(i.e.~$(1,1)$ vortices), this actually looks like a vortex with
winding number $N+\alpha$ with $\phi$ multivalued when
$\alpha\notin\Z$. 
The singular gauge transformation
\beq
\phi \to \sqrt{\frac{\bar{z}^{\alpha}}{z^{\alpha}}}\phi=\frac{|z|^{\alpha}}{z^{\alpha}}\phi
\eeq
is needed to make this look like a $N$ vortex.\footnote{This will
shove the multivalued $\alpha$ phase of the Higgs field into the gauge
field. }

\item The final case where we return to Liouville's equation and
integrability, is when the defect parameter is a constant. Then we
have Liouville's equation for a surface with Gauss curvature
$\lambda_0-\sigma$. The same scaling argument as invoked in
Ref.~\cite{Manton2} and mentioned above means that we have to consider
only three cases, $\lambda_0-\sigma=-1,0,1$. This means that a
constant defect enables integrable vortices to be constructed on a
surface with the ``wrong'' Gauss curvature. This relates Popov,
Jackiw-Pi, and Ambj\o rn-Olesen vortices.\footnote{Of course physically,
it is semantics whether the VEV $\lambda_0$ is physically the VEV or
it has further contributions from other ``fields'', like
$\lambda_0-\sigma$, when $\sigma$ is a constant.}
\end{itemize}

\section{Product gauge groups}\label{sec:product}

In Ref.~\cite{TW} the authors showed how, in the case of Taubes
vortices, the impurity term could be viewed as being due to a heavy
frozen vortex coupled to a different gauge group.
Their method
describes a coupled gauge theory with two scalar fields, charged under
two gauge groups. In particular, one scalar field is in the
bifundamental representation, while the other is in the fundamental
representation of the second gauge group.
This can be generalized to arbitrarily charged scalar fields and
extends to $(\lambda_{0},\lambda)$ vortices as we now show.

We will follow the notation of Ref.~\cite{GEN2020} and consider the
$\U(1)^2$ gauge theory with two Higgs fields $\phi_A$, $A=1,2$, where
the two gauge fields $A_i^a$ belong to $\U(1)_a$, $a=1,2$.
$A$ is the flavor index, $a$ is the gauge group index and finally,
$i=1,2$ is the spatial index of vector fields. 
The charges of the Higgs fields are contained in the charge matrix $Q$,
specified by the gauge covariant derivative
\beq
D_i\phi_A \d x^i = \d \phi_A - \i\sum_{a} Q_{Aa} A^a \phi_A,
\eeq
with $A^a = A_i^a \d x^i$ being the two 1-forms belonging to the gauge
groups $\U(1)_a$, $a=1,2$, respectively.
We will not assume a nonvanishing determinant of the charge matrix, as
was done in Ref.~\cite{GEN2020}, since it will be an unnecessary
assumption in what follows.
If on the other hand, one restricts to the case $\det Q\neq 0$, the
non-integrable system of hyperbolic vortex equations reduces to that studied
in Ref.~\cite{HY}.
Note that we have absorbed the gauge coupling $e$ by changing length
scales in this paper, which in particle physics would make all
elements of $Q$ integers (or for quarks, multiples of $1/6$).
In this paper, we will allow the elements of $Q$ to take on any real
values. 

The field strengths corresponding to the two gauge groups are given by
\beq
F^a = \d A^a = (\p_1 A_2^a - \p_2 A_1^a)\,\d x^1\wedge\d x^2.
\eeq
The static (potential) energy of the theory is
\begin{align}
V = \frac12\int_{M_0}\left[
\frac{1}{\Omega_0^2}\sum_{a} (F_{12}^a)^2
+\frac{2\lambda}{\Omega_0}\sum_{A}|D_i\phi_A|^2 
+\sum_{a}\Big(\lambda\sum_{A}|\phi_A|^2Q_{Aa} - \lambda_0 r_a\Big)^2
\right]\Omega_0\,\d^2x,
\label{eq:V_prod}
\end{align}
where $r_a$, $a=1,2$, are real constants.
The vacuum of the theory is
\beq
\lambda\sum_A|\phi_A|^2Q_{Aa} = \lambda_0r_a,
\eeq
which we require to hold for every $a$, putting certain conditions on
$r_a$; if the $\{r_a\}$'s do not satisfy the above equation, this is
known as supersymmetry breaking (D-term breaking) in physics.
Note that we have left an ambiguity of the sign of the constant terms
in $\lambda_0$, but $r_a\in\mathbb{R}$ are kept as real numbers for
now, which makes it possible to have different magnitudes or signs in
the two resulting vortex equations.

A Bogomol'nyi trick can readily be performed
\begin{align}
V &= \frac12\int_{M_0}\left[
\sum_{a}\left(\frac{F_{12}^a}{\Omega_0} - \lambda_0 r_a
+ \lambda\sum_A |\phi_A|^2Q_{Aa}\right)^2
+ \frac{8\lambda}{\Omega_0}\sum_{A}|D_{\bar{z}}\phi_A|^2
\right]\Omega_0\,\d^2x\non
&\phantom{=\ }
+ 2\pi\lambda_0\sum_a r_a k^a,
\end{align}
where we have defined the magnetic fluxes
\beq
k^a\mathrel{\mathop:}= \frac{1}{2\pi}\int_{M_0} F^a
= \frac{1}{2\pi}\int_{M_0} F_{12}^a\;\d^2x, \qquad a=1,2.
\label{eq:mag_flux}
\eeq
Note that the magnetic fluxes may not be integers in this model.
It is now easy to read off the Bogomol'nyi equations
\begin{align}
D_{\bar{z}}\phi_A &= 0,\\
\frac{1}{\Omega_0}F_{12}^a &= \lambda_0r_a - \lambda\sum_A|\phi_A|^2Q_{Aa}.
\end{align}
When the above equations are satisfied, the Bogomol'nyi bound is
saturated
\beq
V \geq V_{\rm BPS} = 2\pi\lambda_0\sum_a r_ak^a.
\eeq

The first Bogomol'nyi equation can be solved by the gauge fields
\beq
\sum_a Q_{Aa}A_{\bar{z}}^a = -\i\p_{\bar{z}}\log\phi_A,
\label{eq:Azbarsol}
\eeq
from which $\sum_a Q_{Aa}F_{12}^a$ can be calculated, yielding
\beq
-\frac{2}{\Omega_0}\p_z\p_{\bar{z}}\log|\phi_A|^2
= \lambda_0 \sum_a Q_{Aa} r_a
- \lambda\sum_{a,B} Q_{Aa} Q_{Ba}|\phi_B|^2 
- \frac{2\pi}{\Omega_0}\sum_{r=1}^{N_A} \delta(z - Z_r^A),
\eeq
where we have multiplied the second Bogomol'nyi equation by $Q_{Aa}$
and summed over $a$ and $N_A$ is the number of vortices (zeros) in the
$A$th flavor, counted with multiplicity.
The winding number or the vortex number, $N_A$, is given by
the number of zeros in $\phi_A$, counted with multiplicity.

The winding number of the $A$-th flavor of Higgs field, $N_A$, is linearly
related to the fluxes $k^a$ by the charge matrix. This can be seen by
using $\sum_a Q_{Aa}F_{12}^a$ in Eq.~\eqref{eq:mag_flux}, yielding
\begin{equation}
\sum_a Q_{Aa} k^a
= -\frac{1}{\pi}\int_{M_0} \p_z\p_{\bar{z}}\log|\phi_A|^2\;\d^2x
= -\frac{1}{2\pi\i}\oint_{\p M_0} \p_z \log|\phi_A|^2 \d z.
\end{equation}
Using that at the boundary, the Higgs field tends to
$\phi_A\propto z^{N_A}$,
\beq
\sum_a Q_{Aa}k^a = -N_A,
\eeq
we obtain the relation between the magnetic fluxes $k^a$ and the
winding number $N_A$, as promised.
The fluxes $k^a$ correspond to the degrees of the line bundles 
over $M_0$ with connections $A^a$ and smooth sections $\phi_A$.
For a nondiagonal charge
matrix the sections are linearly mixed according to the above
equation.
Note that although the winding numbers, $N_A\in\mathbb{Z}_{>0}$, are
positive integers, which is necessary for the complex scalar fields to
be single valued (and vortices as opposed to anti-vortices), the
magnetic fluxes $k_a$ need not be integers -- even if all the
components of the charge matrix are integers.

Using the standard decomposition $\phi_A=e^{h_A+\i\chi_A}$, we arrive
at 
\beq
-\frac{4}{\Omega_0}\p_z\p_{\bar{z}}h_A
= \lambda_0\sum_a Q_{Aa} r_a
- \lambda\sum_{a,B} Q_{Aa} Q_{Ba} e^{2h_B}
- \frac{2\pi}{\Omega_0}\sum_{r=1}^{N_A} \delta(z - Z_r^A).
\label{eq:general_vortex}
\eeq
Changing variables from $h_A$ to $g_A$ by
\beq
h_A = g_A + \log\frac{1 - \lambda_0 \sum_a Q_{Aa} r_a |z|^2}{2}, 
\label{eq:h_A_Toda}
\eeq
we get
\beq
4\p_z\p_{\bar{z}}g_A
= \lambda\sum_{a,B} Q_{Aa} Q_{Ba} e^{2g_B}
+2\pi\sum_{r=1}^{N_A}\delta(z - Z_r^A),
\label{eq:coupled_liouville_type}
\eeq
provided that
\beq
\sum_a Q_{Aa}r_a = 1,
\label{eq:geom_cond}
\eeq
holds for all $A=1,2$.
This condition is the compatibility condition for the coupled
Liouville-type equations with respect to the background geometry of
the Riemann manifold $M_0$.

Note that by imposing the above geometric compatibility condition,
$h_A$ of eq.~\eqref{eq:h_A_Toda} reduces to
\beq
h_A = g_A + \log\big(\Omega_0^{-\frac12}\big).
\eeq
Explicitly, this means that we only get the coupled vortex equations
\eqref{eq:coupled_liouville_type} for a Riemann surface of constant
Gauss curvature $K_0=-\lambda_0$.

\subsection{Toda system}\label{sec:Toda}

In the particular case where the charge matrix satisfies
\beq
\sum_a Q_{Aa}Q_{Ba} = K_{AB}
=\begin{pmatrix}
2 & -1\\
-1 & 2
\end{pmatrix},
\label{eq:charge_Cartan}
\eeq
where $K_{AB}$ is the $\SU(3)$ Cartan matrix, then the coupled
Liouville-type equations reduce to the integrable Toda system\footnote{Toda equations occur when $K$ is the Cartan matrix of any simple Lie algebra. Thus there will also be Toda equations corresponding to the Cartan matrix for $G_{2}$. }.
There is a 1-parameter family of solutions to this equation, given by
\beq
Q =
\begin{pmatrix}
 \mp'\frac{\sqrt{2-d^2}}{2} \mp' \frac{\pm\sqrt{3}d}{2}
 & -\frac{d}{2} \pm \frac{\sqrt{3}}{2}\sqrt{2-d^2}\\
 \pm'\sqrt{2-d^2} & d
\end{pmatrix}, \qquad d\in\big[-\sqrt{2},\sqrt{2}\big],
\eeq
where $\pm$ and $\pm'$ are two independent signs and $Q Q^{\rm T}=K$.
This solution for the charge matrix has 
\beq
\det Q = \mp'\pm\sqrt{\det K} = \mp'\pm\sqrt{3},
\eeq
for any value of $d\in\big[-\sqrt{2},\sqrt{2}\big]$.
As a specific example, we may choose $d\mathrel{\mathop:}=0$ yielding
\beq
Q = \frac{1}{\sqrt{2}}
\begin{pmatrix}
\mp' 1 & \pm\sqrt{3}\\
\pm' 2 & 0
\end{pmatrix}
= \frac{1}{\sqrt{2}}
\begin{pmatrix}
-1 & \sqrt{3}\\
2 & 0
\end{pmatrix},
\label{eq:Qspecific}
\eeq
where we have chosen the upper signs on the right-hand side.
The Fayet-Iliopoulos parameters 
\beq
\begin{pmatrix}
r_1\\
r_2
\end{pmatrix}
=
\frac12
\begin{pmatrix}
\mp'\pm \sqrt{3}d \pm'\sqrt{2-d^2}\\
\pm\sqrt{3(2-d^2)} + d
\end{pmatrix}
\eeq
are a solution to the geometric compatibility
condition \eqref{eq:geom_cond} and for the charge
matrix \eqref{eq:Qspecific} corresponding to $d\mathrel{\mathop:}=0$, they reduce to 
\beq
\begin{pmatrix}
r_1\\
r_2
\end{pmatrix}
= \frac{1}{\sqrt{2}}
\begin{pmatrix}
\pm'1\\
\pm\sqrt{3}
\end{pmatrix}
= \frac{1}{\sqrt{2}}
\begin{pmatrix}
1\\
\sqrt{3}
\end{pmatrix},
\eeq
where we have chosen the upper signs on the right-hand side.

We thus arrive at the system of equations
\beq
4\p_z\p_{\bar{z}}g_A
= \lambda\sum_B K_{AB} e^{2g_B}
+2\pi\sum_{r=1}^{N_A}\delta(z - Z_r^A),
\label{eq:general_Toda}
\eeq
with $K_{AB}$ given in Eq.~\eqref{eq:charge_Cartan} and $A,B=1,2$.
If we set $\lambda\mathrel{\mathop:}=-1$,
which corresponds to the coupled Popov, coupled Jackiw-Pi or
coupled Ambj\o rn-Olesen vortices, this is exactly the Toda system 
\beq
4\p_z\p_{\bar{z}}g_A
= -\sum_B K_{AB} e^{2g_B}
+2\pi\sum_{r=1}^{N_A}\delta(z - Z_r^A).
\eeq
The Toda system of equations is known to be integrable and for
$K_{AB}$ of Eq.~\eqref{eq:charge_Cartan}, the solution of the above
system of equations is given in terms of the Kostant-Leznov-Saveliev
solution \cite{Kostant:1979qu,Leznov:1979td,Leznov:1979yx}, which is
given by
\beq
\exp(2g_A) = \frac{1}{2}4\p_z\p_{\bar{z}}\log\det\big(M_A^\dag M_A\big),
\eeq
($A$ not summed over) where $M_A$ is a $3$-by-$A$ rectangular matrix.
Since we only have two flavors ($A=1,2$), the matrices $M_A$ in our
case are given by\footnote{
  In the general case of the Toda system with
  a rank $R$ Lie algebra, the matrices are $(R+1)$-by-$A$ rectangular
  matrices given by 
  $M_A \mathrel{\mathop:}=(u,\p_z u,\p_z^2 u,\cdots,\p_z^{A-1} u)$
  and $u$ is given by $(1,f_1(z),f_2(z),\cdots,f_R(z))^{\rm T}$, where
  $A=1,2,\ldots,R$. }
\beq
M_1 = u, \qquad
M_2 =
\begin{pmatrix}
  u, & \p_z u
\end{pmatrix},
\eeq
with $u(z)$ the holomorphic 3-vector
\beq
u\mathrel{\mathop:}=
\begin{pmatrix}
1\\
f_1(z)\\
f_2(z)
\end{pmatrix}.
\eeq
Here the 3-vector is due to the Toda system being the $\SU(3)$
case, which has the Cartan matrix given in
Eq.~\eqref{eq:charge_Cartan}.
The ramification points of $f_{1,2}(z)$ correspond to vortex
positions.

We can generalize the Kostant-Leznov-Saveliev solution to both signs
of $\lambda=\pm 1$, by introducing the sign in the solution as follows
\beq
\exp(2g_A) = -\frac{1}{2\lambda}4\p_z\p_{\bar{z}}\log\det\big(M_A^\dag W_A\big),
\eeq
($A$ not summed over) where $M_A$ and $W_A$ are $3$-by-$A$ rectangular
matrices.
For the two flavors ($A=1,2$), the matrices $M_A$, $W_A$ in our case
are given by\footnote{
  In the general case of the Toda system with
  a rank $R$ Lie algebra, the matrices are $(R+1)$-by-$A$ rectangular
  matrices given by
  $M_A \mathrel{\mathop:}=(u,\p_z u,\p_z^2 u,\cdots,\p_z^{A-1} u)$ and
  $W_A \mathrel{\mathop:}=(v,\p_z v,\p_z^2 v,\cdots,\p_z^{A-1} v)$
  with $u$ given by $(1,f_1(z),f_2(z),\cdots,f_R(z))^{\rm T}$
  and $v$ given by $(1,-\lambda f_1(z),-\lambda f_2(z),\cdots,-\lambda f_R(z))^{\rm T}$, where
  $A=1,2,\ldots,R$. }
\beq
M_1 = u, \qquad
M_2 =
\begin{pmatrix}
  u, & \p_z u
\end{pmatrix}, \qquad
W_1 = v, \qquad
W_2 =
\begin{pmatrix}
  v, & \p_z v
\end{pmatrix},
\eeq
with $u(z)$ and $v(z)$ the holomorphic 3-vectors
\beq
u\mathrel{\mathop:}=
\begin{pmatrix}
1\\
f_1(z)\\
f_2(z)
\end{pmatrix}, \qquad
v\mathrel{\mathop:}=
\begin{pmatrix}
1\\
-\lambda f_1(z)\\
-\lambda f_2(z)
\end{pmatrix}.
\eeq
For $\lambda=-1$, the original Kostant-Leznov-Saveliev solution is
recovered.

What about the $\lambda=0$ case?
Since there is no mixing between the two vortex equations for
$\lambda=0$, the solution is simply the standard integrable Bradlow
vortex solution of Ref.~\cite{Manton2} in each field, as reviewed in
Sec.~\ref{sec:exotic_review}. 

Remarkably, we have succeeded in constructing an integrable (2-by-2)
vortex system of equations for any of the $(\lambda_0,\lambda)$
vortices, provided the geometric compatibility
condition \eqref{eq:geom_cond} is satisfied.
What the condition imposes is to ensure that both vortex flavors can
get the Fayet-Iliopoulos constant absorbed by the \emph{same}
background geometry.

\subsection{Impurities from product gauge groups}\label{sec:impurity_from_product}
Returning to the general product gauge group setup.
Let us do a partial Bogomol'nyi trick on the energy
functional \eqref{eq:V_prod}, say on the second scalar field $\phi_2$
and the second gauge group $\U(1)_2$:
\begin{align}
V = \frac12\int_{M_0}\bigg[&
\bigg(\frac{F_{12}^2}{\Omega_0} - \lambda_0 r_2 + \lambda\sum_A|\phi_A|^2 Q_{A2}\bigg)^2
+\frac{8\lambda}{\Omega_0}|D_{\bar{z}}\phi_2|^2
+2\lambda_0 r_2 \frac{F_{12}^2}{\Omega_0} \non&
-\frac{2\lambda}{\Omega_0}F_{12}^2|\phi_1|^2 Q_{12}
+\frac{2\lambda}{\Omega_0}F_{12}^1|\phi_2|^2 Q_{21} \non&
+\frac{1}{\Omega_0^2}\big(F_{12}^1\big)^2
+\bigg(-\lambda_0r_1 + \lambda\sum_A|\phi_A|^2 Q_{A1}\bigg)^2
+\frac{2\lambda}{\Omega_0}|D_i\phi_1|^2
\bigg]\Omega_0\;\d^2x,
\end{align}
where the third line contains the terms not included in the
Bogomol'nyi trick and the first line looks like a normal Bogomol'nyi
completion. However, due to the non-diagonal charge matrix, there is a
mismatch of two terms on the second line.
So far, everything is symmetric, but to facilitate the discussion of the
magnetic-type impurity interpretation of Ref.~\cite{TW}, we set
\beq
Q_{21} \mathrel{\mathop:}= 0,
\eeq
which reduces the above energy functional to 
\begin{align}
V = \frac12\int_{M_0}\!\bigg[&
\bigg(\frac{F_{12}^2}{\Omega_0} - \lambda_0 r_2 + \lambda|\phi_1|^2 Q_{12} + \lambda|\phi_2|^2 Q_{22}\bigg)^2
+\frac{8\lambda}{\Omega_0}\big|(\p_{\bar{z}} - \i Q_{22} A_{\bar{z}}^2)\phi_2\big|^2
+2\lambda_0 r_2 \frac{F_{12}^2}{\Omega_0} \non&
-\frac{2\lambda}{\Omega_0}F_{12}^2|\phi_1|^2 Q_{12}\non&
+\frac{1}{\Omega_0^2}\big(F_{12}^1\big)^2
+\big(\lambda_0r_1 - \lambda|\phi_1|^2 Q_{11}\big)^2
+\frac{2\lambda}{\Omega_0}\big|(\p_i - \i Q_{11} A_i^1 - \i Q_{12} A_i^2)\phi_1\big|^2
\bigg]\Omega_0\,\d^2x.
\label{eq:freezing_step}
\end{align}
We will now assume that the two vortex equations, corresponding to the
two first positive definite terms, are satisfied and thus set them to
zero.
The second term, corresponding to $D_{\bar{z}}\phi_2=0$, determines
$\phi_2$ in terms of the gauge field $A^2$.
However, the vortex equation corresponding to the first term is
coupled to $\phi_1$, so the magnetic flux $\frac{F_{12}^2}{\Omega_0}$ depends
on $\phi_1$ and hence so does $\phi_2$.
The last term on the first line is a topological invariant (counting
the number of zeros in the field $\phi_2$) and is hence a constant in the
energy.

It will now be convenient to redefine (diagonalize) the gauge field
for $\phi_1$ as
\beq
Q_{11}\calA \mathrel{\mathop:}= Q_{11}A^1 + Q_{12}A^2 \qquad \Rightarrow\qquad
Q_{11}\calF = Q_{11} F^1 + Q_{12} F^2.
\eeq
Eliminating $F_{12}^1$ and imposing the vortex equations, we
arrive at
\begin{align}
V &= \frac12\int_{M_0}\!\bigg[
\frac{1}{\Omega_0^2}(\calF_{12})^2
+\frac{2\lambda}{\Omega_0}\big|D_i^\calA\phi_1\big|^2
+\big(\lambda_0r_1 - \lambda|\phi_1|^2 Q_{11} - \sigma(x)\big)^2
+\frac{2}{\Omega_0}\sigma(x)\calF_{12}
\bigg]\Omega_0\,\d^2x\non
&\phantom{=\ }
+2\pi\lambda_0\left(r_2 - \frac{Q_{12}}{Q_{11}}r_1\right) k^2,
\end{align}
where we have defined
\begin{align}
D_i^{\calA} &\mathrel{\mathop:}= \p_i - \i Q_{11} \calA_i,\\
\sigma(x) &\mathrel{\mathop:}= -\frac{Q_{12}F_{12}^2}{Q_{11}\Omega_0},
\end{align}
with $k^2$ the topological degree of the vortex background
(i.e.~number of vortices in $\phi_2$ multiplied by $-1/Q_{22}$).

The Bogomol'nyi trick on the first flavor of Higgs fields now leads to
the vortex equations with impurities following Ref.~\cite{TW}
\begin{align}
V = \frac12\int_{M_0}\bigg[
\left(\frac{\calF_{12}}{\Omega_0} - \lambda_0r_1 + \lambda|\phi_1|^2Q_{11} - \sigma(x)\right)^2
+ \frac{8\lambda}{\Omega_0}\big|D_{\bar{z}}^\calA\phi_1\big|^2\bigg]\Omega_0\,\d^2x
+ 2\pi\lambda_0 r_1 k,
\end{align}
where we have dropped the constant $\propto k^2$ and defined
\beq
k\mathrel{\mathop:}= \frac{1}{2\pi}\int_{M_0} \calF
= \frac{1}{2\pi}\int_{M_0} \calF_{12}\;\d^2x.
\eeq
The Bogomol'nyi bound is thus
\beq
V \geq V_{\rm BPS} = 2\pi\lambda_0 r_1 k.
\eeq
We have thus generalized the Tong-Wong construction \cite{TW} of
coupled vortex equations as an interpretation of the magnetic
impurities, to the full five exotic $(\lambda_0,\lambda)$ vortex
equations of Ref.~\cite{Manton2} and additionally with generic charge
matrix $Q$ of the form
\beq
Q_{Aa} = 
\begin{pmatrix}
Q_{11} & Q_{12}\\
0 & Q_{22}
\end{pmatrix}_{Aa},
\eeq
where the rows are flavors of Higgs field and the columns are the gauge
groups.
The imposition of the vortex equations in Eq.~\eqref{eq:freezing_step}
is thought of as freezing one species of vortices and this is what
gives rise to the impurities for the effective theory with only the
other vortex species left.

In the special case of
\beq
Q_{Aa} =
\begin{pmatrix}
1 & -1\\
0 & 1
\end{pmatrix},
\label{eq:Q_quiver}
\eeq
the two Higgs fields have the interpretation of a bifundamental field
$(\phi_1)$ and a fundamental field $(\phi_2)$ of the second gauge
group.  

\subsubsection{What became of the topological charge?}

In Sec.~\ref{sec:exotic_liouville}, we discussed the situation of the
impurities being delta functions and seemingly increasing the number
of vortices in the remaining Higgs field (here $\phi_1$).
What is the interpretation in the coupled vortex system?
-- We can now see more clearly what is happening, because the magnetic
flux $k$ is really a linear combination of the degrees of the two
line bundles.
Writing out the Bogomol'nyi mass in terms of the original degrees
$k^{1},k^{2}$, we have
\beq
V_{\rm BPS} = 2\pi\lambda_0 r_1\left(k^1 + \frac{Q_{12}}{Q_{11}}k^2\right).
\eeq
Remembering now that we discarded a constant proportional to $k^2$,
the total Bogomol'nyi mass is actually
\begin{align}
V_{\rm BPS} &= 2\pi\lambda_0 r_1\left(k^1 + \frac{Q_{12}}{Q_{11}}k^2\right)
+ 2\pi\lambda_0\left(r_2 - \frac{Q_{12}}{Q_{11}} r_1\right)k^2 \non
&= 2\pi\lambda_0 (r_1 k^1 + r_2 k^2),
\end{align}
and we are back to the total energy being the sum of the degrees of
each fiber. The deceiving appearances of an extra
topological charge from the impurity are simply due to
the change of variables. 
We recall that the degrees $k^a$ are not necessarily integers for
arbitrary charge matrices.

\subsection{General coupled vortex equations}\label{sec:general_coupled}

If we consider the general case of the vortex
equations \eqref{eq:general_vortex} without imposing the geometric
compatibility condition \eqref{eq:geom_cond}, the system of equations
is not integrable.
In this section, we consider a specific family of examples and
calculate numerical solutions.
Here we keep $(\lambda_0,\lambda)$ general, but fix the charge matrix
to that of Eq.~\eqref{eq:Q_quiver}, for which the system of vortex
equations reduces to
\begin{align}
-\frac{4}{\Omega_0}\p_z\p_{\bar{z}}h_1 &=
\lambda_0(r_1-r_2)
-\lambda\left(2e^{2h_1} - e^{2h_2}\right)
-\frac{2\pi}{\Omega_0}\sum_{r=1}^{N_1}\delta(z - Z_r^1),\\
-\frac{4}{\Omega_0}\p_z\p_{\bar{z}}h_2 &=
\lambda_0 r_2
-\lambda\left(-e^{2h_1} + e^{2h_2}\right)
-\frac{2\pi}{\Omega_0}\sum_{r=1}^{N_2}\delta(z - Z_r^2).
\end{align}
Choosing the constants to be $r_1=r_2=1$, we arrive at
\begin{align}
-\frac{4}{\Omega_0}\p_z\p_{\bar{z}}h_1 &=
-\lambda\left(2e^{2h_1} - e^{2h_2}\right)
-\frac{2\pi}{\Omega_0}\sum_{r=1}^{N_1}\delta(z - Z_r^1),\label{eq:h1}\\
-\frac{4}{\Omega_0}\p_z\p_{\bar{z}}h_2 &=
\lambda_0
-\lambda\left(-e^{2h_1} + e^{2h_2}\right)
-\frac{2\pi}{\Omega_0}\sum_{r=1}^{N_2}\delta(z - Z_r^2).\label{eq:h2}
\end{align}

\begin{figure}[!htp]
\begin{center}
\includegraphics[width=\linewidth]{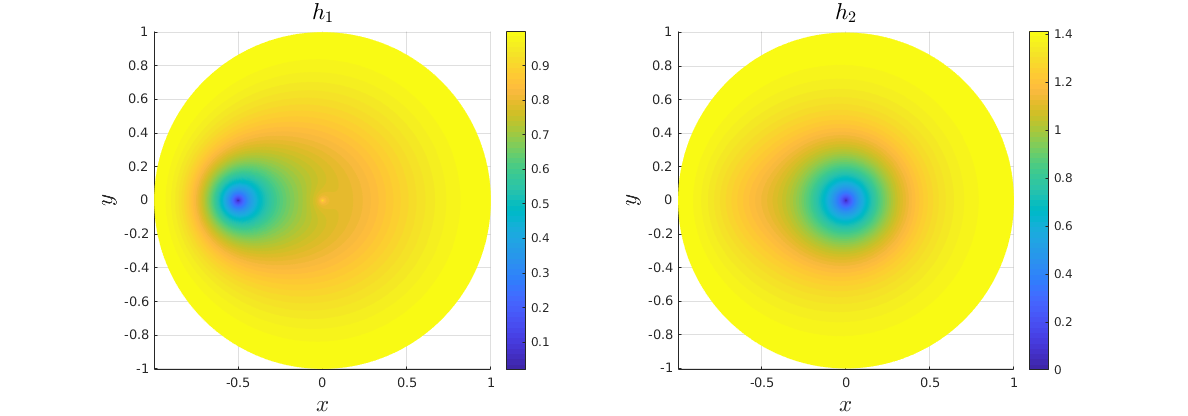}
\caption{A numerical solution of the $(1,1)$ coupled vortices on the
Poincar\'e disk model of $\H^2$. The vortex centers are given in
Eq.~\eqref{eq:vtx_centers}.
The left-hand side panel, showing the field $h_1$, can be interpreted
as a vortex in the presence of an impurity sitting at the center of the
disk.}
\label{fig:coupeldtaubes}
\end{center}
\end{figure}
As an example, we calculate a numerical solution to
Eqs.~\eqref{eq:h1}-\eqref{eq:h2} in the $(1,1)$ case, corresponding to
the coupled Taubes equations with the following vortex positions
\beq
Z_1^1 = -\frac12, \qquad Z_1^2 = 0.
\label{eq:vtx_centers}
\eeq
The result is shown in Fig.~\ref{fig:coupeldtaubes}.
In the context of freezing the second vortex field $h_2$, as was done
in Sec.~\ref{sec:impurity_from_product}, the first vortex $h_1$ can be
interpreted as a single vortex experiencing an impurity sitting at the
origin of the disk.

\section{Discussion and conclusion}\label{sec:discussion}

In this paper, we have considered the five ``exotic'' vortex equations
studied by Manton in Ref.~\cite{Manton2} and generalized all of them
to include magnetic impurities, first put forward for Taubes equations
on $\R^2$ by Tong and Wong in Ref.~\cite{TW}.
The integrability properties considered in Ref.~\cite{CKM} in the case
of Taubes $(\lambda_0,\lambda)=(1,1)$ vortices carry over to all
five exotic vortex equations.
In particular, a delta function impurity does not break integrability
and looks like a source of extra vorticity.
This has a simple interpretation when the impurity is formulated as a
second vortex field in a product gauge group theory, as was considered
already for Taubes vortices in Ref.~\cite{TW}.
For the product gauge group theory, we furthermore generalize the
theory to encompass a generic charge matrix following
Ref.~\cite{GEN2020}. We do not need to impose that the charge matrix
is invertible, as was assumed for the analysis in Ref.~\cite{GEN2020},
but certain constraints on the Fayet-Iliopoulos parameters must hold
for the vacuum to exist.
First we write down the generic case of $(\lambda_0,\lambda)$ vortices
with generic charges and generic Fayet-Iliopoulos parameters.
In order to arrive at a coupled Liouville-type equation, hoping to
obtain the Toda system of equations, we find a geometric compatibility
condition that is a condition on the charge matrix and the
Fayet-Iliopoulos parameters.
A specific choice of charge matrix reduces the system to the $\SU(3)$
Toda equation, which for $\lambda=-1$ is solved by the renowned
Kostant-Leznov-Saveliev 
solution \cite{Kostant:1979qu,Leznov:1979td,Leznov:1979yx}.
We further generalize this solution to the $\lambda=1$ case by a
suitable modification of the Kostant-Leznov-Saveliev solution.
This solution is to the best of our knowledge new.
The $\lambda=0$ case is solved by the Bradlow solution of
Ref.~\cite{Manton2} in each field.
This gives in total five exotic integrable Toda vortex solutions for
Hyperbolic-Toda, Popov-Toda, Jackiw-Pi-Toda, Ambj\o rn-Olesen-Toda and
Bradlow-Toda equations\footnote{The latter is not really a Toda system
of equations, because $\lambda=0$ eliminates the mixing of the
fields.}.
Then we show that the product gauge group theory generalizes the
argument of Tong and Wong \cite{TW} that a second frozen vortex is
interpreted as an impurity of the first species of vortex; our
generalization extends to all five vortex equations and generic charge
matrices of upper-triangular form.
Finally, we construct a numerical solution in the case of coupled
hyperbolic $(\lambda_0,\lambda)=(1,1)$ vortex equations and the
interpretation of the impurity is visualized. 

Our construction can trivially be extended to $N$ coupled vortex
equations in a $\U(1)^N$ gauge theory. All the results carry over by
simply extending the range of the indices, except for the case of the
impurities.
In that case, one would have to decide how many flavors of Higgs
fields are being frozen and how many remain.
Such extension should be straightforward.

One could further consider if there are some natural geometric
interpretations that could be made for some aspects of our
construction, for instance related to the impurities. 
That is, do the integrable cases of the coupled vortex equations admit
a Baptista type geometric interpretation where the vortices give rise
to a degenerate metric \cite{Baptista}? 
Another interesting question is whether there exist further
possibilities of integrable solutions involving an impurity in either
of the vortex equations.

\subsection*{Acknowledgments}

S.~B.~G.~thanks the Outstanding Talent Program of Henan University for
partial support.
The work of S.~B.~G.~is supported by the National Natural Science
Foundation of China (Grants No.~11675223 and No.~12071111).\\
C.~R.~thanks Steffen Krusch for introducing him to the topic of vortices with magnetic impurities.

\end{document}